\documentclass{sig-alternate}

\usepackage{times}
\usepackage{graphicx}
\usepackage{color}
\usepackage[algoruled,vlined]{algorithm2e}
\usepackage{amsmath,amsfonts,amssymb}
\usepackage{listings}
\usepackage{tabularx}
\usepackage{multicol,zi4}
\usepackage{float}
\usepackage{courier,cite}

\usepackage[
  pageanchor=true,
  plainpages=false,
  pdfpagelabels,
  bookmarks,
  bookmarksnumbered,
  pdfborder=0 0 0,  %removes outlines around hyper links in online display
]{hyperref}

\usepackage[T1]{fontenc}

\usepackage[protrusion=true,expansion=true]{microtype}
\usepackage{subfig}

\graphicspath{{figures/}}

\newcommand{\system}{Velox\xspace}

\newcommand{\vmodel}{\texttt{VeloxModel}\xspace}

\newcommand{\ie}{{i.e.,}~}
\newcommand{\eg}{{e.g.,}~}

\begingroup
    \makeatletter
    \@for\theoremstyle:=definition,remark,plain\do{%
        \expandafter\g@addto@macro\csname th@\theoremstyle\endcsname{%
            \addtolength\thm@preskip\parskip
            }%
        }
\endgroup

\floatstyle{ruled}
\newfloat{program}{thp}{lop}
\floatname{program}{Program}

\lstdefinelanguage{scala}{
  morekeywords={abstract,case,catch,class,def,%
    do,else,extends,false,final,finally,%
    for,if,implicit,import,match,mixin,%
    new,null,object,override,package,%
    private,protected,requires,return,sealed,%
    super,this,throw,trait,true,try,%
    type,val,var,while,with,yield},
  otherkeywords={=>,<-,<\%,<:,>:,\#,@},
  sensitive=true,
  morecomment=[l]{//},
  morecomment=[n]{/*}{*/},
  morestring=[b]",
  morestring=[b]',
  morestring=[b]""",
  frame=tb, captionpos=b,
  basicstyle=\small\fontfamily{pcr}\selectfont
}

\newcommand{\minihead}[1]{{\vspace{.4em}\noindent\textbf{#1:} }}
\newcommand{\miniheadit}[1]{{\vspace{.3em}\noindent\textit{#1:} }}

\newcommand{\figref}[1]{Figure~\ref{#1}}
\newcommand{\listref}[1]{Listing~\ref{#1}}

\newcommand{\eqnref}[2]{Eq.~(\ref{#1})\xspace}
\newcommand{\secref}[1]{Section~\ref{#1}}

\newcommand{\reals}{\mathbb{R}}

\newcommand{\size}[1]{\left| #1 \right|}

\begin{document}

% \title{The Missing Piece in Complex Analytics: Low
%   Latency, Scalable Model Deployment and Maintenance with \system }

\title{The Missing Piece in Complex Analytics: Low
  Latency, Scalable Model Management and Serving with \system }

% Prevent email addresses from overflowing hbox -Ankur
\font\eaddfnt=phvr at 11pt \relax

% \numberofauthors{6} %  in this sample file, there are a *total*
% % of EIGHT authors. SIX appear on the 'first-page' (for formatting
% % reasons) and the remaining two appear in the \additionalauthors section.
% \author{
% \alignauthor Daniel Crankshaw \\
%        \affaddr{UC Berkeley}\\
%        \email{crankshaw@cs.berkeley.edu}
% \alignauthor Peter Bailis \\
%        \affaddr{UC Berkeley}\\
%        \email{pbailis@cs.berkeley.edu}
% \alignauthor Joseph E. Gonzalez \\
%        \affaddr{UC Berkeley}\\
%        \email{jegonzal@cs.berkeley.edu}
% \and
% \alignauthor Zhao Zhang\\
%        \affaddr{UC Berkeley}\\
%        \email{zhaozhang@cs.berkeley.edu}
% \alignauthor Haoyuan Li\\
%        \affaddr{UC Berkeley}\\
%        \email{haoyuan@cs.berkeley.edu}
% \alignauthor Michael J. Franklin \\
%        \affaddr{UC Berkeley}\\
%        \email{franklin@cs.berkeley.edu}
% }

\toappear{This article is published under a Creative Commons Attribution License (\href{http://creativecommons.org/licenses/by/3.0/}{http://creativecommons.org/licenses/by/3.0/}), which permits distribution and reproduction in any medium as well allowing derivative works, provided that you attribute the original work to the authors and CIDR 2015.\\[2mm]7th Biennial Conference on Innovative Data Systems Research (CIDR '15)\\January 4-7, 2015, Asilomar, California, USA.}

\author{
Daniel Crankshaw, Peter Bailis, Joseph E. Gonzalez, Haoyuan Li,\\[1mm]Zhao
Zhang, Michael J. Franklin, Ali Ghodsi, Michael I. Jordan\\[2mm]
\affaddr{UC Berkeley AMPLab}
}

\renewcommand*{\ttdefault}{zi4}

\maketitle
\begin{abstract}

  To enable complex data-intensive applications such as personalized
  recommendations, targeted advertising, and intelligent services, the data management community has focused heavily on the design of systems to train complex models on large datasets. Unfortunately, the design
  of these systems largely ignores a critical component of the overall
  analytics process: the serving and management of models at scale.
  In this work, we present \system, a new component of the Berkeley Data
  Analytics Stack. \system is a data management system for facilitating
  the next steps in real-world, large-scale analytics pipelines: online
  model management, maintenance, and serving.
  %, management, and maintenance.
  %{\color{red} HY: consider separating this into two sentences}
  \system provides end-user applications and services with a low-latency,
  intuitive interface to models, transforming the raw statistical
  models currently trained using existing offline large-scale compute
  frameworks into full-blown, end-to-end data products capable of
  targeting advertisements, recommending products, and personalizing
  web content. To provide up-to-date results for these complex models,
  \system also facilitates lightweight online model maintenance and
  selection (i.e., dynamic weighting). In this paper, we describe the
  challenges and architectural considerations required to achieve this
  functionality, including the abilities to span online and offline
  systems, to adaptively adjust model materialization strategies, and
  to exploit inherent statistical properties such as model error
  tolerance, all while operating at ``Big Data'' scale.

\end{abstract}

% \category{H.4}{Information Systems Applications}{Miscellaneous}
%A category including the fourth, optional field follows...
% \category{D.2.8}{Software Engineering}{Metrics}[complexity measures, performance measures]

% \terms{}

% \keywords{ACM proceedings, \LaTeX, text tagging}
%!TEX root = veloxms.tex

\section{Introduction}

The rise of large-scale commodity cluster compute frameworks has enabled the
increased use of complex analytics tasks at unprecedented scale. A large subset
of these complex tasks, which we call \textit{model training} tasks, facilitate
the production of statistical models that can be used to make predictions about
the world, as in applications such as personalized recommendations,
targeted advertising, and intelligent services. By providing scalable platforms
for high-volume data analysis, systems such as Hadoop~\cite{fnt-mapreduce} and
Spark~\cite{spark} have created a valuable ecosystem of distributed model
training processes that were previously confined to an analyst's R console or
otherwise relegated to proprietary data-parallel warehousing engines. The
database and broader systems community has expended considerable energy
designing, implementing, and optimizing these frameworks, both in academia and
industry.

\begin{figure}[t]
\centering
\includegraphics[width=0.8\linewidth]{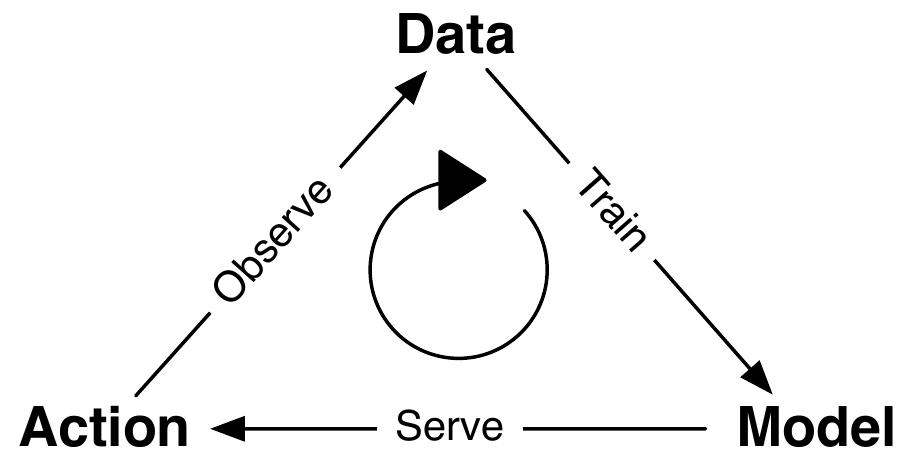}
\vspace{1em} % yes, this has an effect
\caption{\textbf{\system Machine Learning Lifecycle}
  \footnotesize{Velox manages the ML model lifecycle, from model
    training on raw data (the focus of many existing ``Big Data''
    analytics training systems) to the actual actions and predictions
    that the models inform. Actions produce additional observations
    that, in turn, lead to further model training. Velox facilitates
    this cycle by providing functionality to support the
    missing components in existing analytics stacks.}}
\label{fig:mllifecycle}
\vspace{-1em}
\end{figure}

This otherwise productive focus on model training has overlooked a
critical component of real-world analytics pipelines: namely, how are
trained models actually deployed, served, and managed? Consider the
implementation of a collaborative filtering model to recommend songs
for an online music service.  We could use a data-parallel modeling
interface, such as MLbase~\cite{mlbase} on Spark~\cite{spark}, to
build a model of user preferences (say, in the form of a matrix
representing predictions for user-item pairs) based on historical
data---a batch-oriented task for which Spark is well suited and
optimized.  However, if we wish to actually \textit{use} this model to
deliver predictions on demand (e.g., as part of a web service) on
interactive timescales, a data-parallel compute framework such as
Spark is the wrong solution.  Batch-oriented designs sacrifice latency
for throughput, while the mid-query fault tolerance guarantees
provided by modern cluster compute frameworks are overkill and too
costly for fine-grained jobs. Instead, the overwhelming trend in
industry is to simply dump computed models into general-purpose data
stores that have no knowledge of the model semantics. The role of
interpreting and serving models is relegated to another set of
application-level services, and the management of the machine learning
life cycle (\figref{fig:mllifecycle}) is performed by yet another
separate control procedure tasked with model refresh and maintenance.

% The role of interpreting and
% serving models is relegated to another set of application-level services, and
% the entire \emph{machine learning lifecycle} (\figref{fig:mllifecycle}) managed by yet another separate control procedure
% tasked with model refresh and maintenance.

%The past several years have seen calls \cite{Deshpande06, Akdere11,
%Feng12} to introduce various aspects of predictive modeling into existing
%RDBMSs; we argue that, instead of shoehorning model management into traditional
%database engines, the correct approach is to embrace the considerable
%engineering and technical innovations that have enabled these ``Big Learning''
%problems at scale---namely, cluster compute frameworks. 

As a data management community, it is time to address this missing piece in
complex analytics pipelines: machine learning model management and serving at scale.
% \zhao{also in real time?}.
Towards this end, we present \system, a model management platform
within the Berkeley Data Analytics Stack (BDAS). In a sense, BDAS is
prototypical of the real-world data pipelines above: prior to \system,
BDAS contained a data storage manager~\cite{tachyon}, a dataflow
execution engine~\cite{spark}, a stream processor, a sampling engine,
and various advanced analytics packages~\cite{mli}. However, BDAS
lacked any means of actually serving this data to end-users, and the
many industrial users of the stack (e.g., Yahoo!, Baidu, Alibaba,
Quantifind) rolled their own solutions to model serving and
management. \system fills this gap.

Specifically, \system provides end-user applications with a
low-latency, intuitive interface to models at scale, transforming the
raw statistical models computed in Spark into full-blown, end-to-end
data products. Given a description of the statistical model expressed
as a Spark UDF, \system performs two key tasks. First, \system exposes
the model as a service through a generic model serving API providing
low latency predictions for a range of important query types. Second,
\system keeps the models up-to-date by implementing a range of both
offline and online incremental maintenance strategies that leverage
both advances in large-scale cluster compute frameworks as well as
online and bandit-based learning algorithms.

In the remainder of this paper we present the design of the Velox
system and present observations from our initial, pre-alpha
implementation. In \secref{sec:background}, we describe the key
challenges in the design of data products and outline how \system
addresses each. In \secref{sec:architecture}, we provide an overview
of the \system system, including the serving, maintenance, and
modeling components. We discuss our current design and implementation
of each in Sections~\ref{sec:maintenance} through
\ref{sec:modelapi}. Finally, we survey related work in
\secref{sec:related} and conclude with a discussion of ongoing and
future work in \secref{sec:conclusion}.

% Velox integrates with the other technologies in the BDAS stack.
% The Spark cluster compute framework and MLlib are used for fast large-scale offline model training.
% To exchange data and parameters efficiently across the BDAS stack and store large parts of the model state for efficient random access, Velox relies on Tachyon.

% By providing a standard interface to a range of modeling tasks, Velox
% abstracts the complexities of the underlying models.
% Through tight integration with Spark, Velox enables efficient offline retraining of complex models while incrementally maintaining models between batch updates.

% By exploiting bandit methods velx
% training
% frameworks to efficiently publish models that were trained offline to
% a distributed online serving tier.  By grouping models into ensembles,
% Velox leverages advances in online learning to adaptively
% re-weight models on the fly improving robustness and addressing some
% of the challenges associated with introducing new models in an active
% serving environment.

% For a common class of modeling techniques found in recommender systems, Velox introduces new hybrid offline and online learning strategy that leverages the power of cluster compute frameworks to extract overall patterns and then continuously and incremental updates models based on individual user feedback.
% By managing the history of model evaluations, Velox tracks model performance and trigger offline retraining.

%!TEX root = veloxms.tex

\section{Background and Motivation}
\label{sec:background}
\label{sec:matrixfactorization}

% VIGNETTE AND DETAILS ABOUT HOW RECOMMENDATION ACTUALLY WORKS
% A common class of data-products are recommender systems.
% Recommender systems recommend content, ads, and products to individual users by combining behavioral data (\eg ratings and clicks) with data describing users (\eg demographics), the items being considered (\eg content descriptions) and the context (\eg search queries).

%% I wanted to define data products since we use it throughout the paper
Many of today's large scale complex analytics tasks are performed in
service of \textit{data products}: applications powered by a
combination of machine learning and large amounts of input data. These
data products are used in a diverse array of settings ranging from
targeting ads
% %and content to targeting telescopes at vanishing supernovae
and blocking fraudulent transactions to personalized search, real-time
automated language translation, and digital assistants.

\minihead{An example application} To illustrate the implications of data product design on data
management infrastructure, and, as a running example of a data product,
consider the task of building a service to suggest songs to users. This
music suggestion service is an example of a \emph{recommender system},
a popular class of data products which target content to individuals
based on implicit (\eg clicks) or explicit (\eg ratings) feedback.

To begin making recommendations, we start with a \emph{training
dataset} that contains an existing set of user ratings for songs (\eg
songs that interns have manually labeled) and fit a statistical
model to the training set. After this initial training, we iteratively
improve the quality of the model as we collect more ratings.

There are numerous techniques for modeling this data; in this example,
we consider widely used matrix factorization models~\cite{Koren09}. At
a high level, these models embed users and songs into a
high-dimensional space of latent factors and use a distance metric
between each as a proxy for similarity. More formally, these models
\emph{learn} a $d$-dimensional latent factor $w_u \in \reals^d$ for
each user $u$ (collected into a matrix $W \in
\reals^{\size{\text{users}}\times d}$) and $x_i \in \reals^d$ for each song
$i$ (and corresponding $X \in \reals^{\size{\text{songs}}\times d}$).  These
latent factors encode information about unmeasured properties of the
user (\eg \emph{DeadHead}) and song (\eg \emph{PartyAnthem}) and are
learned by solving the following optimization problem:
\[
  \vspace{-0.2em}
\arg \min_{W,X} \lambda \left(||W||_2^2 + ||X||_2^2 \right) + \sum_{(u,i) \in \text{Obs}}
  ( r_{ui} - w_u^T x_i )^2
  \vspace{-0.0em}
\]
Given the $W$ and $X$ matrices, we can calculate a user $u$'s expected
rating for a song $i$ by appropriately projecting $W$ and $X$ to yield
$u$'s weights $w_u$ and $i$'s weights $x_i$, and taking their dot
product:
\[
  \vspace{-0.2em}
  \text{rating}(u,i) = w_u^T x_i
  % \vspace{-0.1em}
\]
Therefore, an implementation of our data product has two natural
phases. The first calculates the optimal $W$ and $X$ matrices
containing factors for each user and item. The second uses $W$ and $X$
to make predictions for specific user-item pairs.

Similar to many companies providing actual data products, we could
implement our song recommendation service by combining cluster compute
frameworks with traditional data management and serving systems.
For the training phase, we might compute $W$ and $X$ periodically (\eg
daily) using a large-scale cluster compute framework like Spark
or GraphLab~\cite{graphlab} based on a snapshot of the ratings logs
stored in a distributed filesystem like HDFS~\cite{hdfs}. In this
architecture, models are trained on stale data and not
updated in response to new user feedback until the next day.

For the serving phase, there are several options. The simplest strategy
would precompute all predictions for every user and song
combination and load these predictions into a lower-latency data store.
While this approach hides the modeling complexity from the serving tier, it has
the disadvantage of materializing potentially billions of predictions
when only a small fraction will likely be required.
%Alternatively, we could load the latent factors in to a data management system and compute the predictions online in a separate prediction service.
Alternatively, a more sophisticated approach might load the latent
factors in to a data management system and compute the predictions
online in the application tier. Given the current lack of a
general-purpose, scalable prediction service, this is likely to be an
ad-hoc task that will be repeated for each data product.

% Unfortunately, this existing infrastructure leads to several challenges, which we outline below and briefly describe design principles that...

% the only ML that goes here is to explain HOW this example works -- add forward reference

% SET OF CHALLENGES
% for each, the problem with existing systems and the approach we take
% Challenge goes here: foo bar baz
% Velox solution: we win!

% Velox is a system for serving, incrementally retraining, and managing machine learning models that addresses the above challenges.

% More generally Velox is a system for hosting a new class of services broadly referred to as data products.
% Data products are applications and services that combine machine learning with vast amounts of data to actively respond to inputs and make decisions that influence others services and products.
% Data products are used in a diverse array of settings ranging from
% targeting ads
% %and content to targeting telescopes at vanishing supernovae
% and blocking fraudulent transactions to personalized search, real-time automated language translation, and digital assistants.
% Because they combine machine learning and with real-time model serving and updates, data products present several unique challenges that differentiate them from both traditional offline analytics systems and online serving systems.

\subsection{Challenges and Opportunities}

While the above straw-man proposal is a reasonable representation of
how users today implement data products, there are several
opportunities for improving the model management experience. Here, we
highlight the challenges in managing a data product that are not
addressed by either traditional offline analytics systems or online
data management systems.

% There are several key challenges to this straw man approach to implementing a data product such as a recommendation service.

\minihead{Low Latency} Because many data products are consumed by
user-facing applications it is essential that they respond within the
window of interactivity to prediction queries and quickly learn from
new information. For example, a user listening to an online radio
station expects their feedback to influence the next songs played by
the music service.
% Alternatively, a fraud detection systems must make predictions about fraud and adapt to changing user habits and context (\eg traveling abroad).
These demands for real-time responsiveness both in their ability to
make predictions and learn from feedback are not well addressed by
traditional cluster compute frameworks designed for scalable but
batch-oriented machine learning. Additionally, while data management
systems provide low latency query serving, they are not capable of
retraining the underlying models.

\miniheadit{Velox's approach} Velox provides low latency query serving by
intelligently caching computation, scaling out model prediction and
online training, and introducing new strategies for fast incremental
model updates.

\minihead{Large scale} With access to large amounts of data, the machine
learning community has developed increasingly sophisticated techniques
capable of modeling data at the granularity of individual entities.
In our example recommender service, we learn factor representations for
individual users and songs.
Furthermore, these latent factor representations are interdependent: changes in the song factors affects the user factors.
% For example, recommender systems often model the interests of individuals as well as the latent characteristics of the items being recommended.
% As a consequence the model state can span millions of records.
% Similarly, deep learning architectures, which have enjoyed recent
% attention by achieving landmark results in a range of important tasks,
% can have billions of parameters that must be evaluated in order to make
% predictions.
The size and interdependency of these models poses unique challenges to our
ability to serve and maintain these models in a low latency
environment.
%\peter{this is weak} \joey{hmmm.}

\miniheadit{Velox's approach} Velox addresses the challenge of
scalable learning by leveraging existing cluster compute frameworks to
initialize the model training process and infer global properties
offline and then applies incremental maintenance strategies to
efficiently update the model as more data is observed. To serve models
at scale, Velox leverages distributed in memory storage and
computation.

% These often non-parametric models grow with the size of the data as
% well as the complexity of the underlying modeled phenomena. As a
% consequence models are no longer concise abstractions of large amounts
% of data but instead data

% Life cycle managment
% catalog of models
% Metrics and regression analysis

\minihead{Model lifecycle management} Managing multiple models and identifying when
models are underperforming or need to be retrained is a key challenge
in the design of effective data products.
%Data products typically have multiple models, feature sets, and require continuous monitoring and retraining.
For example, an advertising service may run a series of ad campaigns,
each with separate models over the same set of users.
%Alternatively, feature engineering may yield new feature transformations that improve prediction accuracy for a collection of models.
Alternatively, existing models may no longer adequately reflect the present
state of the world.
For example, a recommendation system that favors the songs in the Top 40 playlists may become increasingly less
effective as new songs are introduced and the contents of the Top 40
evolves.
Being able to identify when models need to be retrained, coordinating offline training, and updating the online models is essential to provide accurate predictions.
Existing solutions typically bind together separate monitoring
and management services with scripts to trigger retraining, often in an
ad-hoc manner.

%\peter{Do we want to talk about the coordination between batch and serving.}
% \joey{Fixed?}

\miniheadit{Velox's approach} Velox maintains statistics
about model performance and version histories, enabling easier
diagnostics of model quality regression and simple rollbacks to earlier
model versions. In addition, Velox uses these statistics to
automatically trigger offline retraining of models that are under
performing and migrates those changes to the online serving environment.
% and can identify when features functions are no longer performing well.

% Velox addresses these challenges by making user level modeling a first-
% class concept and relying on linear combinations of feature functions
% to support the automatic integration of new models and features.

% However deploying a new model has certain risks that might affect both the user experience as well as the performance of serving system.
% Similarly existing models may no longer adequately reflect the present state of the world.
% For example, a recommendation system that favors the Olympics may become increasingly less effective after the Olympics ends.
%
% Finally in some applications we might introduce user or product
% specific models which capture the behaviors of individuals but also
% share knowledge across individuals.

\minihead{Adaptive feedback} Because data products influence the actions
of other services and ultimately the data that is collected, their
decisions can affect their ability to learn in the future. For example,
a recommendation system that only recommends sports articles may not
collect enough information to learn about a user's preferences for
articles on politics.
%Furthermore, the bias introduced in the data by existing models can adversely affect the training of new models.
While there are a range of techniques in the machine learning
literature~\cite{Watkins89,Li10} to address this feedback loop, these techniques must be able to
modify the predictions served and observe the results, and thus run
in the serving layer.
% . Because of this,
% these algorithms run in
% need to run
% at the query interface as they dynamically adapt the model
% predictions.\peter{This is confusing: what does it mean to run at the
% query interface?}
%are only well studied in the context of linear models and must be run within the query interface.

\miniheadit{Velox's approach} Velox adopts bandit techniques~\cite{Li10} for
controlling feedback that enable the system to optimize not only
prediction accuracy but its ability to effectively learn the user
models. % (\secref{subsec:bandits}).

%!TEX root = veloxms.tex

\section{System Architecture}
\label{sec:architecture}

In response to the challenges presented in Section~\ref{sec:background}, we
developed \system, a system for serving, incrementally maintaining, and managing
machine learning models at scale within the existing BDAS ecosystem. \system
allows BDAS users to build, maintain, and operate full, end-to-end data
products. In this section, we outline the \system architecture.
%  which we examine in detail in the remainder of this paper.

\system consists of two primary architectural components.
First, the \textbf{\system model manager} orchestrates the computation and
maintenance of a set of pre-declared machine learning models, incorporating
feedback and new data, evaluating model performance, and retraining models as
necessary. The manager performs fine-grained model updates but, to facilitate
large scale model re-training, uses Spark for efficient batch computation.

% In effect, the model manager acts as a combined system catalog and workflow
% manager.

Second, the \textbf{\system model predictor} implements a low-latency,
up-to-date prediction interface for end-users and other data product consumers.
There are a range of pre-materialization strategies for model predictions, which
we outline in detail in \secref{sec:serving}.

To persist models and training data,
\system uses a configurable storage backend. By default,
\system is configured to use Tachyon~\cite{tachyon}, a fault-tolerant,
memory-optimized distributed storage system in BDAS. In our
current architecture, both the model manager and predictor are
deployed as a pair of co-located processes that are resident with each
Tachyon worker process. When coupled with an intelligent
routing policy, this design maximizes data locality.

\begin{figure}[t]
\centering
\includegraphics[width=\linewidth]{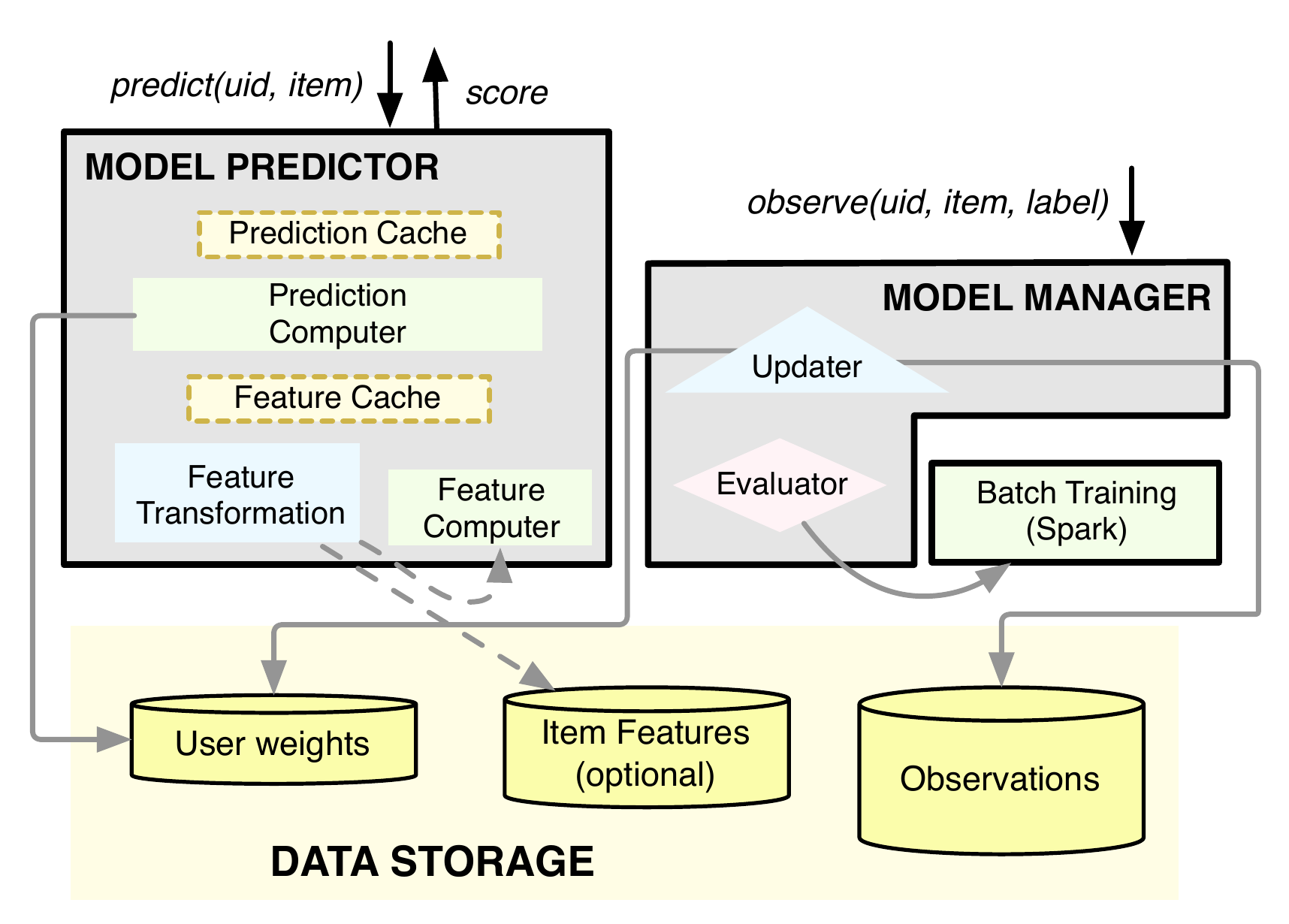}
%\vspace{0pt} % yes, this has an effect
\caption{\textbf{\system System Architecture}
    \footnotesize{Two core components, the
    Model Predictor and Model Manager, orchestrate low latency and
    large-scale serving of data products computed (originally, in
    batch) by Spark and stored in a lightweight storage layer (e.g., Tachyon).}
}
\label{fig:sys-architecture}
%\vspace{-1em}
\end{figure}

% \subsection {Machine Learning Models in \system}
% \label{subsec:machinelearning}

\minihead{Current modeling functionality} The current \system
implementation provides native support for a simple yet highly
expressive family of personalized linear models that generalizes the
matrix factorization model presented in
\secref{sec:matrixfactorization}. (We discuss extensions in
Section~\ref{sec:conclusion}.)  Each model consists of a
$d$-dimensional weight vector $w_u \in \mathbb{R}^d$ for each user
$u$, and a feature function $f$ which maps an input object (\eg a
song) into a $d$-dimensional feature space (\eg its latent factor
representation). A prediction is computed by evaluating:
% the following linear function:
\begin{equation}
%\vspace{-0.4em}
\text{prediction}(u, x) = w_u^T f(x, \theta) \label{eqn:glm}
%\vspace{-0.2em}
\end{equation}
The feature parameters $\theta$ in conjunction with the feature
function $f$ enable this simple model to incorporate a wide range of
models including support vector machines (SVMs)~\cite{Burges98}, deep
neural networks~\cite{Bengio09}, and the latent factor models used to
build our song recommendation service.

% By using SVMs or neural networks as features we can express complex non-linear models.
% Alternatively by using the parameters $\theta$ to store the entire latent feature representation of the songs in our music recommendation service we can express the latent feature models central to modern recommendation services.

\minihead{Recommendation service behavior} In our music recommendation data
product, the \system prediction service (\secref{sec:serving}) computes
predictions for a given user and item (or set of items) by reading the user
model $w_u$ and feature parameters $\theta$ from Tachyon and evaluating
\eqnref{eqn:glm}~. The \system model maintenance service
(\secref{sec:maintenance}) updates the user-models $w_u$ as new observations
enter the system and evaluates the resulting model quality. When the model
quality is determined to have degraded below some threshold, the maintenance
service triggers Spark, the offline training component, to retrain the feature
parameters $\theta$. Spark consumes newly observed data from the storage layer,
recomputes the user models and feature parameters, and writes the results back
to Tachyon.

In the subsequent sections, we provide greater detail about the design of each
component, the interfaces they expose, and some of the design decisions we have
made in our early prototype. While our focus is on exposing these generalized
linear models as data products to end-users, we describe the process of
implementing additional models in Section~\ref{sec:modelapi}.

\section{MODEL MANAGEMENT}
\label{sec:maintenance}

The model management component of \system is responsible for
orchestrating the model life-cycle including: collecting observation
and model performance feedback, incrementally updating the user
specific weights, continuous evaluation of model performance, and the
offline retraining of feature parameters.

% new observations coming in, plus API

% online maintenance

% bulk recomputation
% \subsection{Model Creation}
%
% Models are created by using the following simple api...
%
% Feature functions may be created by using some more complex api defined later...
%
%
% \textbf{Warm Start:}  One of the key challenges in predictive services is how to model new users.
% Here we adopt a simple heuristic in which new users are assigned a global $w_u$ which was computed based on the last offline update.
% If $g$ is a linear function we can improve upon this heuristic by assigning new users the recent estimate of the average of the existing user weight vectors which yields predictions corresponding to the average prediction for all users.

\subsection{Feedback and Data Collection}

As users interact with applications backed by \system, the front-end
applications can gather more data (both explicitly and implicitly) about a user's
behavior. \system exposes an \emph{observation} interface to consume this new
interaction data and update the user's model accordingly.
% The observation interface is independent of the prediction interface, although often an observation about a user's preference for an item will arrive shortly after a prediction recommending that item to the user.
To insert an observation about a user-item pair into \system, a front-end
application calls \texttt{observe} (\listref{listing:predictapi}),
% \begin{center}
% \begin{lstlisting}[language=scala, label=listing:observeapi]
% def observe(uid: Long, x: Data, y: double)
% \end{lstlisting}
% \end{center}
providing the user's ID, the item data (for feature extraction),
and the correct label $y$ for that item.
% To evaluate the quality of the updated user-model, the maintenenance service needs the
% label that the old model would have predicted, determined by requesting a prediction
% prior to the model update.
In addition to being used to trigger online updates, the observation is written
to Tachyon for use by Spark when retraining the model offline.

% \begin{figure}[t]
% \begin{lstlisting}[language=scala, label=listing:observeapi, caption={\textbf{The Observation API}
% \vspace{-1em}
% }]
% def observe(uid: Long, x: Data, y: double)
% \end{lstlisting}
% \end{figure}

\subsection{Offline + Online Learning}

Learning in the \system modeling framework consists of estimating the user specific weights $w_u$ as well as the parameters $\theta$ of the feature function.
To simultaneously enable low latency learning and personalization while also supporting sophisticated models we divide the learning process into online and offline phases.
The offline phase adjusts the feature parameters $\theta$ and can run infrequently because the feature parameters capture aggregate properties of the data and therefore evolve slowly.
The online phase exploits the independence of the user weights and the linear structure of \eqnref{eqn:glm}~ to permit lightweight conflict free per user updates.

The infrequent offline retraining phase leverages the bulk computation capabilities of large-scale cluster compute frameworks to recompute the feature parameters $\theta$.
The training procedure for computing the new feature parameters is defined as an opaque Spark UDF and depends on the current user weights and all the available training data.
The result of offline training are new feature parameters as well as potentially updated user weights.

Because the offline phase modifies both the feature parameters and user weights it invalidates both prediction and feature caches.
To alleviate some of the performance degradation resulting from invalidating both caches, the batch analytics system also computes all predictions and feature transformations that were cached at the time the batch computation was triggered.
These are used to repopulate the caches when switching to the newly trained model.
We intend to investigate further improvements in this process, as it is possible
that the set of hot items may change as the retrained models redistribute the
distribution of popularity among items. % Great idea!

\begin{figure}[t]
\includegraphics[width=\linewidth]{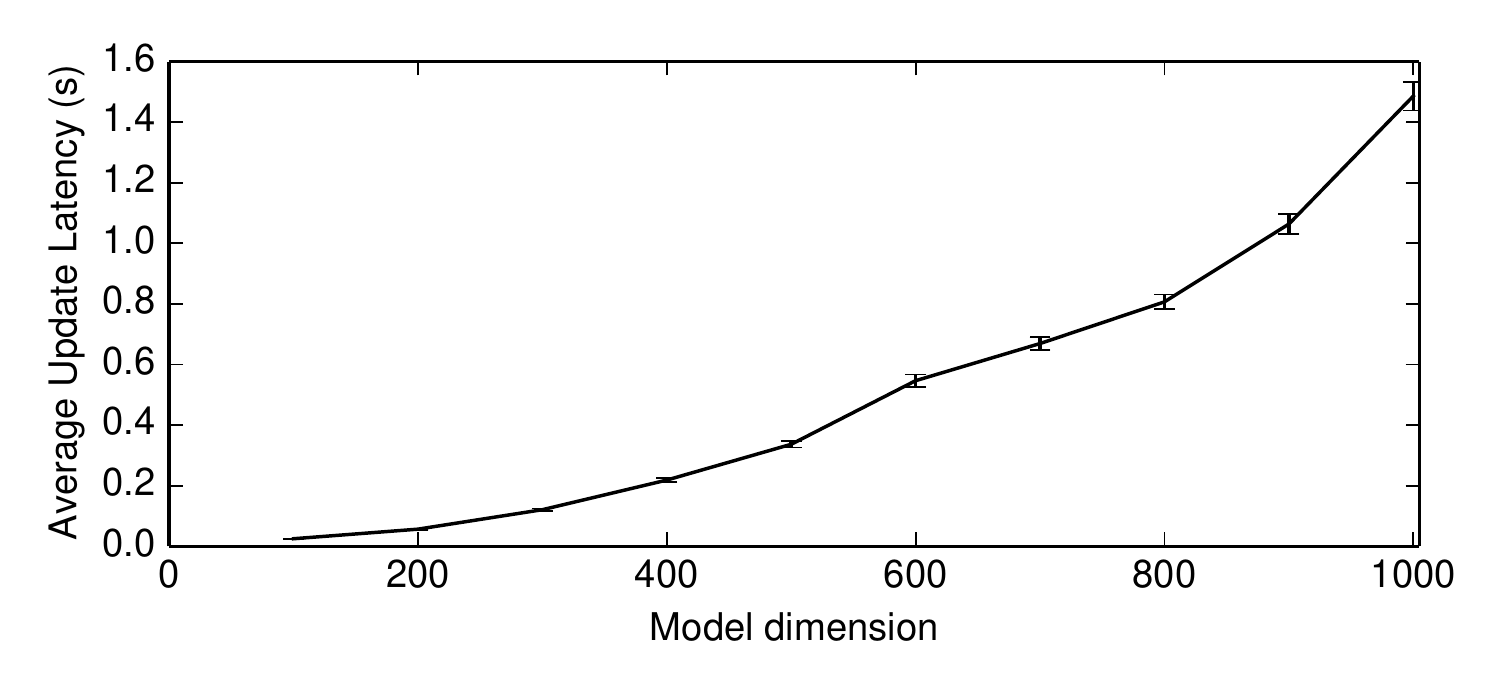}
\vspace{-1em} % yes, this has an effect
\caption{\textbf{Update latency vs model complexity}
  \footnotesize{Average time to perform an online update to a user
  model as a function of the number of factors in the model. The results are averaged
  over 5000 updates of randomly selected users and items from the MovieLens 10M rating data
  set. Error bars represent 95\% confidence intervals.}
%\vspace{-1.5em}
}
\vspace{0.5em}
\label{fig:update-latency}
\end{figure}

The online learning phase runs continuously and adapts the user specific models $w_u$ to observations as they arrive.
While the precise form of the user model updates depends on the choice of error function  (a configuration option) we restrict our attention to the widely used squared error (with $L_2$ regularization) in our initial prototype.
As a consequence the updated value of $w_u$ can be obtained analytically using the normal equations:
\begin{equation}
w_u \leftarrow (F(X, \theta)^T F(X, \theta) + \lambda I_n)^{-1} F(X, \theta)^T Y
  \vspace{-0.3em}
\end{equation}
where $F(X, \theta) \in \reals^{n_u \times d}$ is the matrix formed by evaluating the feature function on all $n_u$ examples of training data obtained for that particular user and $\lambda$ is a fixed regularization constant.
While this step has cubic time complexity in the feature dimension $d$ and linear time complexity in the number of examples $n$ it can be maintained in time quadratic in $d$ using the Sherman-Morrison formula for rank-one updates.
Nonetheless using a naive implementation we are able to achieve (\figref{fig:update-latency}) acceptable latency for a range of feature dimensions $d$ on a real-world collaborative filtering task.

By updating the user weights online and the feature parameters
offline, we are provide an \textit{approximation} to continuously retraining the entire model.
Moreover, while the feature parameters evolve slowly, they still change. By not continuously updating their value, we potentially introduce inaccuracy into the model.
To assess the impact of the hybrid online + offline incremental strategy adopted by Velox, we evaluated the accuracy of Velox on the MovieLens10M
dataset\footnote{\url{http://grouplens.org/datasets/movielens}}.
By initializing the latent features with 10 ratings from each user and then using an additional 7 ratings, we were able to achieve 1.6\% improvement in prediction accuracy\footnote{Differences in accuracy on the MovieLens dataset are typically measured in small percentages.} by applying the online strategy.
This is comparable to the 2.3\% increase in accuracy achieved using full offline retraining.

%If we restrict our attention to new users, those that have no existing rating, we are able to achieve a 22.8\% increase in accuracy by obtaining just 7 ratings with and applying the incremental training algorithm.

We first used offline training to initialize the feature parameters $\theta$ on half of the data and then evaluated the prediction error of the proposed strategy on the remaining data.
By using the \system's incremental online updates to train on 70\% of the remaining data, we were able to achieve a held out prediction error that is only slightly worse than complete retraining.

% \begin{figure}[t]
% \includegraphics[width=\linewidth]{data/model-quality}
% \vspace{0pt} % yes, this has an effect
% \caption{
% \textbf{The Root Mean Squared Error (RMSE) of various incremental update strategies.} We compare the error (lower is better) associated with not retraining, retraining using the Velox fast incremental update strategy, and retraining the entire model offline.
% \vspace{-1em}
% }
% \label{fig:accuracy}
% \end{figure}

\subsection{Model Evaluation}

Monitoring model performance is an essential part of any predictive service.
\system relies on model performance metrics to both aid administrators in managing deployed models and to identify when offline retraining of feature parameters is required.
To assess model performance, \system applies several strategies.
First, \system maintains running per-user aggregates of errors associated with each model.
Second, \system runs an additional cross-validation step during incremental user weight updates to assess generalization performance.
Finally, when the \texttt{topK} prediction API is used, \system employs bandit algorithms to collect a pool of validation data that is not influenced by the model.
When the error rate on any of these metrics exceeds a pre-configured threshold, the model is retrained offline.

% By tracking the overall model performance Velox can trigger the offline learning phase to execute in the background to update potentially out-of-date feature models (see \secref{sec:modelapi} for details).

%!TEX root = veloxms.tex

\section{Online Prediction Service}

\label{subsec:servepredictions}

%\section{Serving}
\label{sec:serving}

% Split into two tasks: prediction

% detailed prediction API

% how are predictions made?

% GLM stuff here

\begin{figure}[t]
  \begin{lstlisting}[language=scala, label=listing:predictapi, caption={\textbf{The Prediction and Observation API} \footnotesize{These methods form the front-end API for a prediction
      and model management service (\ie \system).}
\vspace{2em}
}]
def predict(s: ModelSchema, uid: UUID, x: Data)
  : (Data, Double)
def topK(s: ModelSchema, uid: UUID, x: List[Data])
  : List[(Data, Double)]
def observe(uid: UUID, x: Data, y: Double)
\end{lstlisting}
\vspace{-3.5em}
\end{figure}

The Velox prediction service exposes model predictions to other
services and applications through a simple interface
(\listref{listing:predictapi}) The \texttt{predict} function serves
point predictions for the provided user and item, returning the item
and its predicted score. The \texttt{topK} function evaluates the best
item from the provided set for the given \texttt{uid}. Support for
the latter function is necessary for \system to implement the bandit
methods described later in this section and can be used to support pre-filtering
items according to application level policies.

\minihead{Caching} The dominant expense when serving predictions in \system is evaluating
the feature function $f$. In particular, when $f$ represents a
materialized feature function (\eg matrix factorization models), the
distributed lookup of the latent factor in $\theta$ is the dominant
cost. Alternatively, when $f$ represents a computational feature
function (\eg a deep neural network) the computation becomes the
dominant cost. These costs reflect two opportunities for
optimization: caching the results of feature function evaluations and
efficiently partitioning and replicating the materialized feature
tables to limit remote data accesses. \system performs both caching
strategies in the Velox predictor process, corresponding to the
\emph{Feature Cache} in \figref{fig:sys-architecture}. In addition, we
can cache the final prediction for a given (user,item) pair, often
useful for repeated calls to \texttt{topK} with overlapping itemsets,
corresponding to the \emph{Prediction Cache} in
\figref{fig:sys-architecture}.

\begin{figure}[t]
\centering
\includegraphics[width=\linewidth]{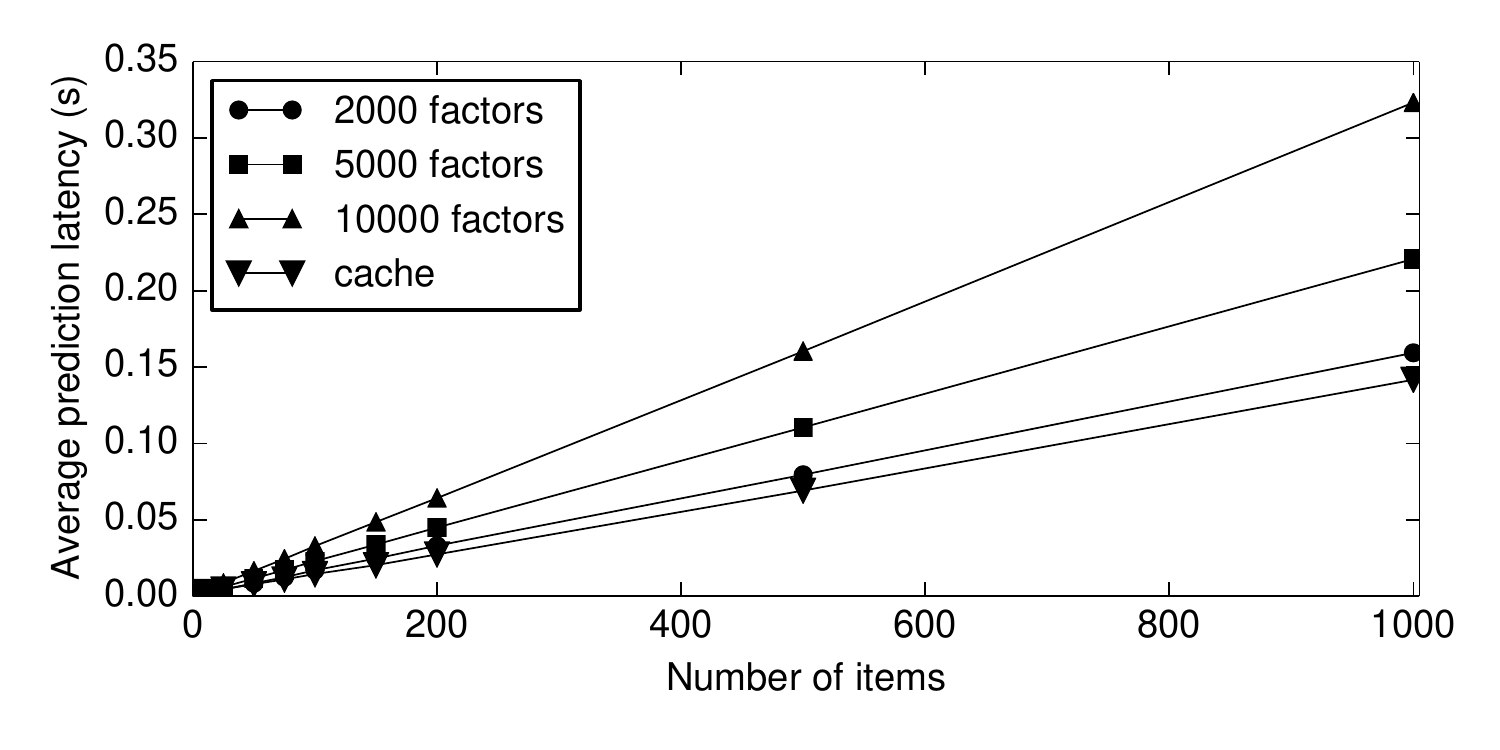}
\vspace{-1em} % yes, this has an effect
\caption{\textbf{Prediction latency vs model complexity}
  \footnotesize{Single-node \texttt{topK} prediction latency for both cached and non-cached
    predictions for the Movie Lens 10M rating dataset, varying size of
    input set and dimension ($d$, or, factor).  Results are averaged over 10,000 trials.}
  %\vspace{-1em}
  }
\label{fig:predict-latency}
\end{figure}

To demonstrate the performance improvement that the prediction cache
provides, we evaluated the prediction latency of computing
\texttt{topK} for itemsets of varying size. We compare the prediction
latency for the optimal case, when all predictions are cached (\ie
100\% cache hit rate) with the prediction latencies for models of
several different sizes. As \figref{fig:predict-latency} demonstrates,
the relationship between itemset size and prediction latency grows
linearly, which is to be expected. And as the model size grows (a
simple proxy for the expense of computing a prediction, which is a
product of both the prediction expense and the feature transformation
expense), the benefits of caching grow.

To distribute the load across a \system cluster and reduce network
data transfer, we exploit the fact that every prediction is associated
with a specific user and partition $W$, the user weight vectors table,
by \texttt{uid}. We then deploy a routing protocol for incoming user
requests to ensure that they are served by the node containing that
user's model. This partitioning serves a dual purpose. It ensures
that lookups into $W$ can always be satisfied locally, and it provides
a natural load-balancing scheme for distributing both serving load and
the computational cost of online updates. This also has the beneficial
side-effect that all writes --- online updates to user weight vectors
--- are local.

When the feature function $f$ is materialized as a pre-computed
lookup table, the table is also partitioned across the
cluster. Therefore, evaluating $f$ may involve a data transfer
from a remote machine containing the required item-features pair.
However, while the number of items in the system may be large, item
popularity often follows a Zipfian distribution~\cite{Meka09}.
Consequently, many items are not frequently accessed, and a small
subset of items are accessed very often. This suggests that caching
the hot items on each machine using a simple cache eviction strategy
like LRU, will tend to have a high hit rate. Fortunately, because the
materialized features for each item are only updated during the
offline batch retraining, cached items are invalidated infrequently.
%never updated and thus never need to be invalidated.

When the feature transformation $f$ is computational, caching
the results of computing the basis functions can provide similar
benefits. For popular items, caching feature function evaluation
reduces prediction latency by eliminating the time to compute the
potentially expensive feature function, and reduces computational load
on the serving machine, freeing resources for serving queries.

\minihead{Bootstrapping} One of the key challenges in predictive
services is how to model new users. In \system, we currently adopt a
simple heuristic in which new users are assigned a recent estimate of
the average of the existing user weight vectors:
\[
\bar{w}^T f(x, \theta) = \frac{1}{|\text{users}|} \sum_{u \in \text{users}} w_u^T f(x, \theta)
  \vspace{-0.4em}
\]
This also corresponds predicting the average score for all users.
%% Do we know this is true ->
%, a strategy that has performed well in practice thus far.

\minihead{Bandits and Multiple Models} Model serving influences
decisions that may, in turn, be used to train future models. This
can lead to feedback loops.  For example, a music recommendation
service that only plays the current Top40 songs will never receive
feedback from users indicating that others songs are preferable. To
escape these feedback loops we rely on a form of the \emph{contextual
 bandits algorithm}~\cite{Li10}, a family of techniques developed to
avoid these feedback loops. These techniques assign each item an
\emph{uncertainty} score in addition to its
predicted store. The algorithm improves models greedily by reducing
uncertainty about predictions. To reduce the total uncertainty
in the model, the algorithm recommends the item with the best
\emph{potential} prediction score (\ie the item with max sum of score
and uncertainty) as opposed to recommending the item with the absolute
best prediction score. When \system observes the correct score for
that recommendation and an online update is triggered, that update
will reduce the uncertainty in the user weight vector more
so than an observation about an item with less uncertainty. The
bandits algorithms exploit the \texttt{topK} interface to select the
item that has the highest potential predicted rating. For example, if
Velox is unsure to what extent a user is a \emph{DeadHead} it will
occasionally select songs such as ``New Potato Caboose'' to evaluate this
hypothesis even if those songs do not have the highest prediction
score.

\section{Adding New Models}
\label{sec:modelapi}

It is possible to express a wide range of models and machine learning techniques within Velox by defining new feature functions.
%As a consequence \system also enables data scientists to implement new models via user defined feature transformations.
To add a new model to \system, a data scientist implements and uploads a new \vmodel instance (\listref{listing:vmodel}). % and uploads it to \system,

\begin{figure}[t]
  \begin{lstlisting}[language=scala, label=listing:vmodel, caption={\textbf{The \texttt{VeloxModel} Interface}
    Developers can add new models and feature transformation functions to \system
    by implementing this interface. Implementations specify how to featurize items,
    perform offline training, and how to evaluate model quality.
    % , the three functions needed by \system to serve predictions from and maintain a model.
\vspace{-1em}
}]
class VeloxModel:
  val name: String  // <- user provided
  val state: Vector // <- feature parameters
  val version: Int  // <- system provided
  def VeloxModel(state: Opt[Vector])
  // Feature Transformation Function
  def features(x: Data, materialized: Boolean)
      : Vector
  // Learning
  def retrain(f: (Data) => Vector,
              w: Table[String, Vector],
              newData: Seq[Data])
      : ((Data) => Vector, Table[String,Vector])
  // Quality Evaluation
  def loss(y: Label, yPredict: Label,
           x: Data, uid: UUID): Double
\end{lstlisting}
%g\vspace{-1em}
\end{figure}

\minihead{Shared state} Each \vmodel may be instantiated with a
vector, used to provide any global, immutable state (\ie $\theta$ from
\secref{sec:background}) needed during the featurization process. For
example, this state may be the parameters for a set of SVMs learned
offline and used as the feature transformation function.

\minihead{Feature transformations} The \vmodel function \texttt{features}
implements the feature transformation function. The \texttt{features} function may
implement a computation on some input data, as is the case when the
feature transformation is the computation of a set of basis
functions. Alternatively, the \texttt{features} function may implement a lookup of the
latent features in a table, similar to the table $W$ used to store the
user models. The implementor indicates which of these two strategies
is used by explicitly specifying whether the features are materialized or are
computed. Continuing with the ensemble of SVMs example, \texttt{features}
would evaluate a set of SVMs with different parameters (stored in the member
\texttt{state}) passed in on instance construction. We are investigating
automatic ways of analyzing data dependencies through techniques like UDF byte-code inspection.

\minihead{Quality evaluation and model retraining} The user provides
two functions, \texttt{retrain} and \texttt{loss}, that allow \system
to automatically detect when models are stale and retrain them. The
\texttt{loss} is evaluated every time new data is observed (\ie every
time a user model is updated) and if the loss starts to increase
faster than a threshold value, the model is detected as stale. Once a
model has been detected as stale, \system retrains the model offline
using the cluster compute framework.  \texttt{retrain} informs the
cluster compute framework how to train this \vmodel, as well as where
to find and how to interpret the observation data needed for
training. When offline training completes, \system automatically
instantiates a new \vmodel and new $W$ --- incrementing the version
--- and transparently upgrades incoming prediction requests to be
served with the newly trained user-models and \vmodel.

% This abstraction has several benefits. By treating all user models as linear weight vectors,
% the system can always perform online updates for users, independent of the schema
% implementation. Specifying whether a feature transformation has data dependencies
% \system can decide on the best caching strategy. And providing support for both computational
% and lookup-based feature transformations allows \system to
% support a wide range of model implementations.

% as an exmaple, discuss ensembles "Say we want to add new feature x"---

% how do we add new models to the system

% talk about the API, feature function, etc.

% \input{system}
%
% \input{results}
%!TEX root = veloxms.tex

\section{Related Work}
\label{sec:related}

\system draws upon a range of related work from the intersection of database
systems and complex analytics. We can broadly characterize this work
as belonging to three major areas:

\minihead{Predictive database systems} The past several years have seen several
calls towards tight coupling of databases and predictive analytics. The Longview
system~\cite{Akdere11} integrates predictive models as first-class entities in
PostgreSQL and introduces a declarative language for model querying. Similarly,
Bismarck~\cite{Feng12} allows users to express complex analytics via common
user-defined aggregates. A large body of work studies probabilistic databases,
which provide first-class support for complex statistical models but, in turn,
focus on modeling uncertainty in data~\cite{Suciu11,Wang08}. Commercially, the
PMML markup language and implementations like
Oryx\footnote{\url{https://github.com/cloudera/oryx}} provide support for a
subset of the data product concerns addressed by \system. Our focus in \system is
to provide predictive analytics as required for modern data products in a large
scale distributed setting. In doing so, we focus on user-specific
personalization, online model training, and the challenges of feedback loops in
modern predictive services.

In contrast with prior work on model management in database systems,
which was largely concerned with managing deterministic metadata (such
as schema)~\cite{bernstein-model}, \system focuses on the use and
management of statistical models from the domain of machine learning.

\minihead{View materialization} The problem of maintaining models at scale can
be viewed as an instance of complex materialized view maintenance. In
particular, MauveDB exploits this connection in a single-node context, providing
a range of materialization strategies for a set of \emph{model-based views}
including regression and Kalman filtering~\cite{Deshpande06} but does not
address latent feature models or personalized modeling. Similarly,
Columbus~\cite{Zhang14} demonstrates the power of caching and model reuse in
in-situ feature learning. We see considerable opportunity in further exploiting the
%impressive amount of
literature on materialized view maintenance~\cite{fnt-matviews} in the model serving
%complex analytics
setting.

\minihead{Distributed machine learning} There are a bevy of systems suitable for
the performing batch-oriented complex analytics tasks~\cite{fnt-mapreduce}, and
a considerable amount of work implementing specific tasks. For example, Li et
al.~\cite{Li13} explored a strategy for implementing a variant of SGD within the
Spark cluster compute framework that could be used by \system to improve offline
training performance. Our focus is on leveraging these existing algorithms to
provide better \textit{online} predictive tasks. However, we aggressively exploit
these systems' batch processing ability and large install bases in our solution.

\section{Conclusions and Future Work}
\label{sec:conclusion}

In this paper, we introduced \system, a system for performing machine learning model serving and model maintenance at scale. \system leverages knowledge of prediction semantics to efficiently cache and replicate models across a cluster. \system updates models to react to changing user patterns, automatically monitoring model quality and delegating offline retraining to existing cluster compute frameworks. In doing so, \system fills a void in current production analytics pipelines, simplifying front-end applications by allowing them to consume predictions from automatically maintained complex statistical models.
% as a service.

We have completed an initial \system prototype that exposes a RESTful client interface and integrates with existing BDAS components, relying on Spark and Tachyon for offline training and distributed data storage.  By running tests against the MovieLens10M dataset we demonstrated that our early prototype performs well on basic serving and model update tasks.  In addition we have evaluated our online incremental update strategy and demonstrated that it closely recovers the prediction accuracy of offline batch retraining.

We are actively pursuing several areas of further research within \system. While we have chosen a fairly conservative modeling interface thus far, we are investigating alternative prediction and modeling APIs---in particular, their effect on more sophisticated query planning and materialization strategies. We plan to integrate and evaluate additional multi-armed bandit (i.e., multiple model) techniques from the machine learning literature (including their dynamic updates) as well as more efficient top-K support for our linear modeling tasks. Velox will
be released as open source software, and we anticipate an alpha code release in early 2015.

\section*{Acknowledgments}
The authors would like to thank Joseph M. Hellerstein, Tomer Kaftan,
Henry Milner, Ion Stoica, Vikram Sreekanti, and the anonymous CIDR
reviewers for their thoughtful feedback on this work.  This research
is supported in part by NSF CISE Expeditions Award CCF-1139158, LBNL
Award 7076018, and DARPA XData Award FA8750-12-2-0331, the NSF
Graduate Research Fellowship (grant DGE-1106400), and gifts from
Amazon Web Services, Google, SAP, The Thomas and Stacey Siebel
Foundation, Adobe, Apple, Inc., Bosch, C3Energy, Cisco, Cloudera, EMC,
Ericsson, Facebook, GameOnTalis, Guavus, HP, Huawei, Intel, Microsoft,
NetApp, Pivotal, Splunk, Virdata, VMware, and Yahoo!.

\newpage

\bibliographystyle{abbrv}

\bibliography{references}

\begin{thebibliography}{10}

\bibitem{Akdere11}
M.~Akdere et~al.
\newblock The case for predictive database systems: Opportunities and
  challenges.
\newblock In {\em CIDR}, 2011.

\bibitem{fnt-mapreduce}
S.~Babu and H.~Herodotou.
\newblock Massively parallel databases and mapreduce systems.
\newblock {\em Foundations and Trends in Databases}, 5(1), 2013.

\bibitem{Bengio09}
Y.~Bengio.
\newblock Learning deep architectures for {AI}.
\newblock {\em Found. Trends Mach. Learn.}, 2(1):1--127, Jan. 2009.

\bibitem{bernstein-model}
P.~A. Bernstein.
\newblock Applying model management to classical meta data problems.
\newblock In {\em CIDR}, 2003.

\bibitem{Burges98}
C.~J.~C. Burges.
\newblock A tutorial on support vector machines for pattern recognition.
\newblock {\em Data Min. Knowl. Discov.}, 2(2):121--167, June 1998.

\bibitem{fnt-matviews}
R.~Chirkova and J.~Yang.
\newblock Materialized views.
\newblock {\em Foundations and Trends in Databases}, 4(4):295--405, 2012.

\bibitem{Deshpande06}
A.~Deshpande and S.~Madden.
\newblock {MauveDB}: Supporting model-based user views in database systems.
\newblock In {\em SIGMOD}, 2006.

\bibitem{Feng12}
X.~Feng, A.~Kumar, B.~Recht, and C.~R{\'e}.
\newblock Towards a unified architecture for in-rdbms analytics.
\newblock In {\em SIGMOD}, 2012.

\bibitem{graphlab}
J.~E. Gonzalez et~al.
\newblock Powergraph: Distributed graph-parallel computation on natural graphs.
\newblock In {\em OSDI}, 2012.

\bibitem{Koren09}
Y.~Koren, R.~Bell, and C.~Volinsky.
\newblock Matrix factorization techniques for recommender systems.
\newblock {\em IEEE Computer}, 42(8):30--37, Aug. 2009.

\bibitem{mlbase}
T.~Kraska, A.~Talwalkar, J.~C. Duchi, R.~Griffith, M.~J. Franklin, and M.~I.
  Jordan.
\newblock Mlbase: A distributed machine-learning system.
\newblock In {\em CIDR}, 2013.

\bibitem{Li13}
B.~Li, S.~Tata, and Y.~Sismanis.
\newblock Sparkler: Supporting large-scale matrix factorization.
\newblock In {\em EDBT}, 2013.

\bibitem{tachyon}
H.~Li, A.~Ghodsi, M.~Zaharia, S.~Shenker, and I.~Stoica.
\newblock Tachyon: Reliable, memory speed storage for cluster computing
  frameworks.
\newblock In {\em SOCC}, 2014.

\bibitem{Li10}
L.~Li, W.~Chu, J.~Langford, and R.~E. Schapire.
\newblock A contextual-bandit approach to personalized news article
  recommendation.
\newblock In {\em WWW}, 2010.

\bibitem{Meka09}
R.~Meka et~al.
\newblock Matrix completion from power-law distributed samples.
\newblock In {\em NIPS}. 2009.

\bibitem{hdfs}
K.~Shvachko et~al.
\newblock The hadoop distributed file system.
\newblock In {\em MSST}. IEEE, 2010.

\bibitem{mli}
E.~R. Sparks et~al.
\newblock {MLI: An API} for distributed machine learning.
\newblock In {\em ICDM}, 2013.

\bibitem{Suciu11}
D.~Suciu, D.~Olteanu, C.~R{\'e}, and C.~Koch.
\newblock {\em Probabilistic Databases}.
\newblock Synthesis Lectures on Data Management. Morgan {\&} Claypool
  Publishers, 2011.

\bibitem{Wang08}
D.~Z. Wang et~al.
\newblock Bayesstore: Managing large, uncertain data repositories with
  probabilistic graphical models.
\newblock 2008.

\bibitem{Watkins89}
C.~Watkins.
\newblock {\em Learning from Delayed Rewards}.
\newblock PhD thesis, King's College, Cambridge, UK, May 1989.

\bibitem{spark}
M.~Zaharia et~al.
\newblock Resilient distributed datasets: A fault-tolerant abstraction for
  in-memory cluster computing.
\newblock In {\em NSDI}, 2012.

\bibitem{Zhang14}
C.~Zhang, A.~Kumar, and C.~R{\'e}.
\newblock {Materialization optimizations for feature selection workloads}.
\newblock In {\em SIGMOD}, 2014.

\end{thebibliography}
% ACM needs 'a single self-contained file'!
%\balancecolumns
% \appendix
%Appendix A
\end{document}